\documentclass{desyproc}
\graphicspath{{figures/}}
\begin{document}
\title{QCD Vacuum as Domain Wall Network}

\author{{\slshape Sergei N. Nedelko$^1$, Vladimir E. Voronin$^1$}\\[1ex]
$^1$Bogoliubov Laboratory of Theoretical Physics, JINR,
141980 Dubna, Russia}
\contribID{xy}

\confID{999}  
\desyproc{DESY-PROC-2099-01}
\acronym{VIP2010} 
\doi  

\maketitle

\begin{abstract}
Domain model of confinement, chiral symmetry breaking and hadronization is based on description of QCD vacuum gluon configurations as an ensemble of almost everywhere homogeneous Abelian (anti-)self-dual fields. The ensemble can be explicitly constructed as domain wall network. We shortly overview this approach and within this framework  discuss the effects of QCD vacuum polarization by strong electromagnetic fields. It is stressed that such polarization effects can play the role of trigger for deconfinement.     
\end{abstract}

\section{Domain wall network}
The following effective Lagrangian for gluon field and its motivation were discussed in~\cite{Galilo:2010fn}:
\begin{gather}
    \mathcal{L}_{\mathrm{eff}} = - \frac{1}{4\Lambda^2}\left(D^{ab}_\nu F^b_{\rho\mu} D^{ac}_\nu F^c_{\rho\mu} + D^{ab}_\mu F^b_{\mu\nu} D^{ac}_\rho F^c_{\rho\nu }\right)  -U_{\mathrm{eff}}   
    \nonumber \\
U_{\mathrm{eff}}=\frac{\Lambda^4}{12} {\rm Tr}\left(C_1\breve{ f}^2 + \frac{4}{3}C_2\breve{ f}^4 - \frac{16}{9}C_3\breve{ f}^6\right). \label{effpot}
\end{gather}
Here $\Lambda$ is a scale of QCD related to gluon condensate, $\breve f=\breve F/\Lambda^2$, and  
\begin{gather*}
D^{ab}_\mu = \delta^{ab} \partial_\mu - i\breve{ A}^{ab}_\mu = \partial_\mu - iA^c_\mu {(T^c)^{ab}},
\\
 F^a_{\mu\nu} = \partial_\mu A^a_\nu - \partial_\nu A^a_\mu - if^{abc} A^b_\mu A^c_\nu,
\quad \breve{ F}_{\mu\nu} = F^a_{\mu\nu} T^a,\ \ \ T^a_{bc} = -if^{abc}
\\
 {\rm Tr}\left(\breve{ F}^2\right) = \breve{ F}^{ab}_{\mu\nu}\breve{ F}^{ba}_{\nu\mu} = -3 F^a_{\mu\nu}F^a_{\mu\nu} \leq 0,\quad C_1>0, \ C_2>0, \ C_3 > 0.
\end{gather*}
The minima of the effective potential are achieved for covariantly constant Abelian (anti-)self-dual fields
\begin{equation*}
\breve{ A}_{\mu}  = -\frac{1}{2}\breve{ n}_k F_{\mu\nu}x_\nu, \, \tilde F_{\mu\nu}=\pm F_{\mu\nu},\quad F_{\mu\nu}F_{\mu\nu}=b_\mathrm{vac}^2\Lambda^4
\end{equation*}
where the matrix $\breve{n}_k$ belongs to the Cartan subalgebra of $su(3)$
\begin{equation*}
 \breve n_k = T^3\ \cos\left(\xi_k\right) + T^8\ \sin\left(\xi_k\right),
\quad
\xi_k=\frac{2k+1}{6}\pi, \, k=0,1,\dots,5.
\end{equation*}
The minima are connected by discrete parity and Weyl transformations, which indicates that the system is prone to existence of solitons in real space-time and kink configurations in Euclidean space. Group-theoretical analysis of symmetry breaking and domain wall formation was done in~\cite{George:2012sb}.

Let us consider the simplest kink configuration -- kink interpolating between self-dual and anti-self-dual Abelian vacua. If the angle $\omega$ between chromoelectric and chromomagnetic fields is allowed to deviate from the constant vacuum value and all other parameters  are fixed to the vacuum values, then the Lagrangian reads
\begin{equation*} 
\mathcal{L}_{\textrm{eff}} = -\frac{1 }{2}\Lambda^2 b_{\textrm{vac}}^2 \partial_\mu \omega \partial_\mu \omega 
- b_{\textrm{vac}}^4  \Lambda^4 \left(C_2+3C_3b_{\textrm{vac}}^2 \right){\sin^2\omega}.
\end{equation*}
The corresponding equation of motion is sine-Gordon equation
\begin{equation*}
    \partial^2\omega = m_\omega^2 \sin 2\omega,\ \   m_\omega^2 = b_{\textrm{vac}}^2  \Lambda^2\left(C_2+3C_3b_{\textrm{vac}}^2 \right),
\end{equation*}
which have the following kink solution:
\begin{equation*}
 \omega(x_\nu) = 2\ {\rm arctg} \left(\exp(\mu x_\nu)\right).
\end{equation*}
The angle $\omega$ interpolates between $0$ and $\pi$.  $x_\mu$ stays for one of the four Euclidean coordinates. The kink describes a planar domain wall between the regions with almost homogeneous Abelian self-dual and anti-self-dual gluon fields. Chromomagnetic and chromoelectric fields are orthogonal to each other on the wall. Far away from the wall the field is self-dual or anti-self-dual, and absolute value of the topological charge density is equal to the value of the gluon condensate. The topological charge density vanishes on the wall.

We may now construct kink superposition with the standard methods. Let us denote the general kink configuration as
\begin{equation*}
 \zeta(\mu_i,\eta_\nu^{i}x_\nu-q^{i})=\frac{2}{\pi}\arctan\exp(\mu_i(\eta_\nu^{i}x_\nu-q^{i})),
\end{equation*}
where $\mu_i$ is the inverse width of the kink, $\eta_\nu^{i}$ is a normal vector to the plane of the wall. The general kink network is then given by
\begin{equation}
\label{kink_network}
\omega=\pi\sum_{j=1}^{\infty}\prod_{i=1}^k \zeta(\mu_{ij},\eta_\nu^{ij}x_\nu-q^{ij}).
\end{equation}

\begin{figure}
\parbox[c][][c]{0.5\textwidth}{
\includegraphics[width=0.5\textwidth]{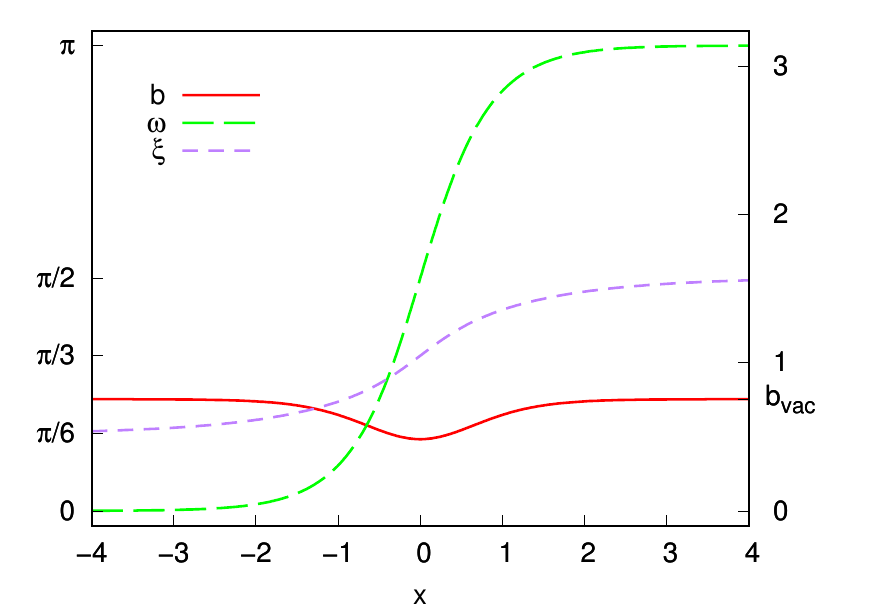}}
\parbox[c][][c]{0.49\textwidth}{
\caption{Kink connecting distinct vacua. Angle $\omega$ varies from $0$ to $\pi$, angle $\xi$ varies from $\pi/6$ to $\pi/2$. Global minima of effective potential~\eqref{effpot} are degenerate in the value of field strength $b=b_\mathrm{vac}$, so its magnitude is the same for all vacua. Dimensionless variable $x\equiv \Lambda x$ is used. \label{3-kink}}}
\end{figure}

Now we consider more general kink-like solution: three parameters $b_\mathrm{vac},\omega$ and $\xi$ deviate from their vacuum values near the center of the kink (Fig.~\ref{3-kink}). One may see that the value of scalar condensate $F_{\mu\nu}F_{\mu\nu}=b_\mathrm{vac}^2\Lambda^4$ decreases in the center of the kink which may increase size of stable chromomagnetic flux tube.

\section{Chromomagnetic trap}
\begin{figure}[h!]
\parbox[t][][c]{0.49\textwidth}{\centering\includegraphics[width=45mm,angle=0]{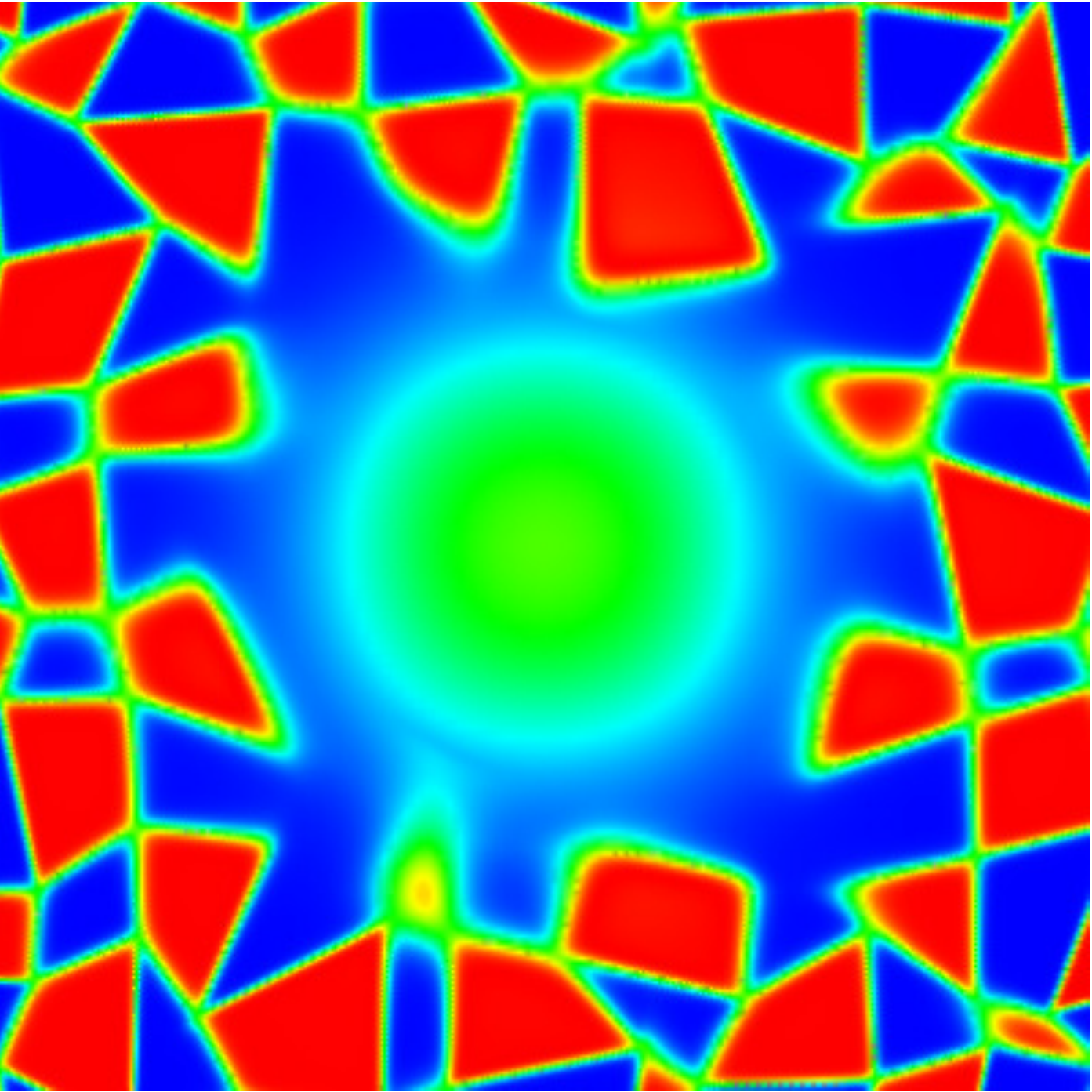}}\hfill
\parbox[t][][c]{0.49\textwidth}{\includegraphics[width=0.49\textwidth,angle=0]{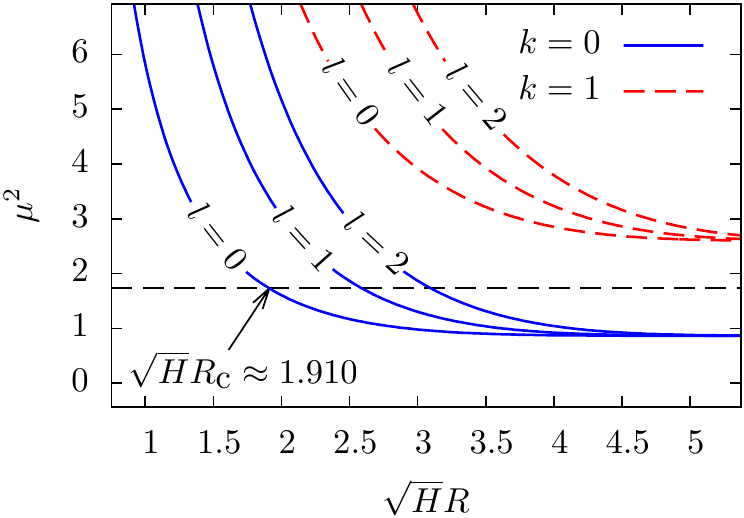}}
\parbox[t][][t]{0.45\textwidth}{\caption{ Example of two-dimensional slice of the cylindrical thick domain wall junction. Self-dual  and anti-self-dual fields are indicated by blue and red colors respectively. The topological charge densityvanishes inside the junction (green color). \label{Fig:chromo_bag}}}\hfill
\parbox[t][][t]{0.45\textwidth}{\caption{The lowest eigenvalues corresponding to positive color orientation $\kappa^a=1$ as functions of $\sqrt{H}R$. The critical radius $R_{\rm c}$ corresponds to $\mu_{a00}^2=2v=\sqrt{3}$. For large $\sqrt{H}R$ eigenvalues  approach correct Landau levels, the degeneracy in $l$ is restored.}}
 \label{Fig:Eigen_flow}
\end{figure}

Calculation of one-loop quark contribution to the QCD effective potential demonstrated that strong electromagnetic fields can play the role of a trigger for  deconfinement~\cite{NG2011-1}. The catalyzing role of strong magnetic fields was also observed in Lattice QCD simulations~\cite{Bonati:2016kxj,Bali:2013owa,Bali:2014kia}. During heavy ion collisions, the electric $\mathbf E_{\rm el}$ and magnetic $\mathbf H_{\rm el}$ fields are almost orthogonal to each other~\cite{Voronyuk:2011jd,Skokov:2009qp}: $\mathbf E_{\rm el}\mathbf H_{\rm el}\approx 0 $. For this configuration of the external electromagnetic field, the one-loop
quark contribution to the QCD effective potential for the
homogeneous Abelian gluon fields is minimal for the  chromoelectric and chromomagnetic fields directed along the electric and magnetic fields respectively~\cite{NG2011-1,Ozaki:2013sfa}.  The orthogonal chromo-fields are not confining: color charged quasiparticles can  move along the chromomagnetic field. Deconfined quarks as well as gluons can move preferably along the direction of the magnetic field even after the electromagnetic field turns off.
This means that a defect in domain wall network may be formed during heavy ion collision (see Fig.~\ref{Fig:chromo_bag}).

Let us consider eigenvalue problem for vector field
\begin{equation}
\label{vector1}
\left[-\breve D^2\delta_{\mu\nu}+2i\breve n B_{\mu\nu}\right]Q_\nu=\lambda^2 Q_\mu
\end{equation}
with the boundary conditions
\begin{equation*} 
\breve n Q_{\mu}(x)=0, \  x\in \partial\mathcal{T},\quad
\partial\mathcal{T} =\left\{x_1^2+x_2^2=R^2, \ (x_3,x_4)\in \mathrm{R^2}  \right\}.
\end{equation*}
The corresponding eigenvalues are
\begin{gather*}
\lambda^2_{alk\nu}=p_4^2+p_3^2+\mu_{alk}^2+ 2s_\nu\kappa_a v,\quad
k=0,1,\dots,\infty, \ \ l\in Z,
\\
s_1=1, \ s_2=-1, \ s_3=s_4=0, \ \kappa_a=\pm 1.
\nonumber
\end{gather*}
In the finite trap the lowest eigenvalue is
\begin{equation*}
\lambda^2_{a00\nu}=p_4^2+p_3^2+\mu_{a00}^2-2v, \  s_\nu\kappa_a=-1.
\end{equation*}
Few lowest eigenvalues $\mu_{akl}^2$ as functions of $\sqrt{H}R$ are shown in Fig.~\ref{Fig:Eigen_flow}. 
One concludes that if the dimensionless size $\sqrt{H}R$ of the trap is sufficiently small 
\begin{eqnarray}
\label{dmlsc}
\sqrt{H}R<\sqrt{H}R_{\rm c}\approx 1.91,
\end{eqnarray}
  then there are no unstable tachyonic modes in the spectrum of color charged vector fields.
  
To estimate the critical size one may use the mean phenomenological value of the gluon condensate (gauge coupling constant $g$ is included into the field strength tensor) 
\begin{equation*}
\langle F^a_{\mu\nu}F^{a\mu\nu}\rangle = 2H^2\approx 0.5{\rm GeV}^4.
\end{equation*} 
Equation~\eqref{dmlsc} leads to the critical radius
\begin{equation*}
R_{\rm c}\approx 0.51 \ {\rm fm} \ (2R_{\rm c}\approx 1 \ {\rm fm}).
\end{equation*}
However, as it follows from Fig.~\ref{3-kink}, scalar condensate become smaller at the domain wall and may increase the value of critical radius.

In Minkowski space-time the eigenvalue problem
turns to the wave equation:
\begin{equation*}
\left[-\breve D^2\delta_{\mu\nu}+2i\breve n B_{\mu\nu}\right]Q_\nu=0.
\end{equation*}
Solution for this equation can be found as a linear combination of eigenmodes~\eqref{vector1} with $\lambda=0$.
\begin{gather}
\label{mass_mink}
p_0^2=p_3^2+\mu_{alk}^2+ 2s_\nu\kappa_a v,
\\ \nonumber
p_0=\pm\omega_{alk\nu}(p_3), \ \omega_{alk\nu}=\sqrt{p_3^2+\mu_{alk\nu}^2+ 2s_\nu,\kappa_a v}.
\end{gather}

Equation \eqref{mass_mink} can be treated as the dispersion relation between energy $p_0$ and momentum $p_3$ for the quasiparticles. Unlike the chromomagnetic field, the (anti-)self-dual fields characteristic for the bulk of domain network configuration lead to purely discrete spectrum of eigenmodes in Euclidean space and do not possess any quasiparticle treatment in terms of  dispersion relation between energy and momentum for elementary  color charged excitations. 

\section{Discussion}
Even though a single chromomagnetic trap has critical size, overall volume of deconfinement transition in hadronic matter is not restricted by $L_c$. Numerous traps may be formed, for instance, due to the influence of strong electromagnetic fields in the whole region of relativistic heavy ion collision~\cite{NG2011-1}.   The traps may merge into a large chromomagnetic  lump developing the  flux tube structure inside the lump. Equivalently, one may also think about formation of the flux tubes  inside the traps with overcritical size emerging under  conditions of high energy or/and baryon number density, strong electromagnetic fields. 

In the absence of any external impact one expects that the fraction of the deconfinement phase is statistically negligible as it occupies an essentially three-dimensional sub-manifold in the four-dimensional space $R^4$.  
The influence of external conditions like the high energy density, high baryon density and strong electromagnetic fields  can trigger the  growth of the relative fraction of the deconfined phase. Under certain conditions high energy density accumulated by colorless collective excitations  can be converted into the thermodynamics of color charged quasiparticles.  The growth of the deconfinement phase fraction  may be described in terms of creation of either numerous  stable chromomagnetic traps or/and the trap with overcritical size, populated by the polarized chromomagnetic tubes~\cite{Ambjorn:1980ms}.   

In the confinement regime (color charge quasiparticles are completely  suppressed) 
one may  expect
$$ q(x)=\langle |g^2\tilde F(x)F(x)|\rangle_{ \cal B}=\langle  g^2F(x)F(x)\rangle_{ \cal B}=B_{\rm vac}^2, $$
while the deconfinement regime (color charged quasiparticles are fully activated) occurs if
$$ q(x)\ll \langle g^2 F(x)F(x)\rangle_{ \cal B}=B_{\rm vac}^2. $$

The scalar gluon condensate may stay practically unchanged during the phase transformation. The regime of dilute plasma can not be reached unless the scalar gluon condensate density vanishes, which would require complete rearrangement of
the quantum effective action. However, as soon as the color charged quasiparticles are activated the structure of the effective action itself
in principle becomes dependent on their temperature and density, and it cannot be excluded that the scalar condensate vanishes at certain critical 
values of the temperature and density.

In summary, the domain wall  network representation of QCD vacuum is suggestive of a two-stage deconfinement transition. At the first stage 
topological charge density vanishes, but the scalar condensate stays almost unchanged, color charged quasiparticles are activated at this stage while
the colorless collective excitations can decay into the color charged ones. The system is far from being a dilute gas. At the second stage the  
scalar gluon condensate vanishes, and the system turns into dilute quark-gluon plasma.

\begin{footnotesize}

\end{footnotesize}

\begin{thebibliography}{99}
\bibitem{Galilo:2010fn} 
  B.~V.~Galilo and S.~N.~Nedelko,
  Phys.\ Part.\ Nucl.\ Lett.\  {\bf 8}, 67 (2011)
 [arXiv:1006.0248 [hep-ph]].

\bibitem{George:2012sb} 
  D.~P.~George, A.~Ram, J.~E.~Thompson and R.~R.~Volkas,
  Phys.\ Rev.\ D {\bf 87}, 105009 (2013)
  [arXiv:1203.1048 [hep-th]].

\bibitem{NG2011-1} B.~V.~Galilo and S.~N.~Nedelko, Phys. Rev. D84 (2011) 094017.

\bibitem{Bonati:2016kxj} 
  C.~Bonati, M.~D'Elia, M.~Mariti, M.~Mesiti, F.~Negro, A.~Rucci and F.~Sanfilippo,
  Phys.\ Rev.\ D {\bf 94}, no. 9, 094007 (2016)
  [arXiv:1607.08160 [hep-lat]].

\bibitem{Bali:2013owa} 
  G.~S.~Bali, F.~Bruckmann, G.~Endrodi and A.~Schafer,
  Phys.\ Rev.\ Lett.\  {\bf 112}, 042301 (2014)
  [arXiv:1311.2559 [hep-lat]].

\bibitem{Bali:2014kia} 
  G.~S.~Bali, F.~Bruckmann, G.~Endr\"odi, S.~D.~Katz and A.~Sch\"afer,
  JHEP {\bf 1408}, 177 (2014)
  [arXiv:1406.0269 [hep-lat]].

\bibitem{Voronyuk:2011jd}
  V.~Voronyuk, V.~D.~Toneev, W.~Cassing, E.~L.~Bratkovskaya, V.~P.~Konchakovski and
S.~A.~Voloshin,
  Phys.\ Rev.\ C {\bf 83}, 054911 (2011)
  [arXiv:1103.4239 [nucl-th]].

\bibitem{Skokov:2009qp}
  V.~Skokov, A.~Y.~Illarionov and V.~Toneev,
  Int.  J.  Mod. Phys.  A {\bf 24} (2009) 5925
  [arXiv:0907.1396 [nucl-th]].

\bibitem{Ozaki:2013sfa} 
  S.~Ozaki,
  Phys.\ Rev.\ D {\bf 89}, no. 5, 054022 (2014)
  [arXiv:1311.3137 [hep-ph]].

\bibitem{Ambjorn:1980ms} 
  J.~Ambjorn and P.~Olesen,
  Nucl.\ Phys.\ B {\bf 170}, 265 (1980).
\end{thebibliography}
\end{document}